\RequirePackage{fix-cm}
\documentclass[smallextended]{svjour3}       
\smartqed  
\usepackage{array}
\usepackage{graphicx}
\usepackage{subfigure}

\usepackage{float}
\usepackage{multirow}
\usepackage{rotating}
\usepackage{epstopdf}
\usepackage{amsmath} 
\usepackage{amssymb}  
\usepackage{color}

\usepackage{latexsym}
 \journalname{Wireless Networks}
\begin{document}

\title{Multi-destination Aggregation with Binary Symmetric Broadcast Channel Based Coding in 802.11 WLANs%
\thanks{Work supported by Science Foundation Ireland grant 07/IN.1/I901 and 11/PI/11771. Both authors were with Hamilton Institute, NUI Maynooth.}
}


\author{Xiaomin Chen         \and
        Douglas Leith 
}


\institute{Xiaomin Chen \at
              Department of Computer and Information Sciences, Northumbria University, UK \\
              Tel.: 00441912274391\\
              \email{xiaomin.chen@northumbria.ac.uk}           
           \and
           Douglas Leith \at
               School of Computer Science and Statistics, Trinity College Dublin, Ireland
}

\date{Received: 08 Aug 2017 / Accepted: 27 Nov 2017}

\maketitle

\begin{abstract}
In this paper we consider the potential benefits of adopting a
binary symmetric broadcast channel paradigm for multi-destination
aggregation in 802.11 WLANs, as opposed to a more conventional
packet erasure channel paradigm. We propose two approaches
for multi-destination aggregation, \emph{i.e.} superposition coding and a simpler time-sharing
coding. Theoretical and simulation results for both unicast and
multicast traffic demonstrate that increases in network throughput
of more than 100\% are possible over a wide range of network
conditions and that the much simpler time-sharing scheme yields
most of these gains and have minimal loss of performance.
Importantly, these performance gains are achieved exclusively
through software rather than hardware changes.
\end{abstract}

\section{Introduction}

Increasing the PHY rates used in a WLAN leads to faster transmission of
the packet payload of a frame, but the overheads associated with each
transmission (PHY header, MAC contention time etc) typically
do not decrease at the same rate and thus begin to dominate the
frame transmission time. To maintain throughput efficiency at
high PHY rates, 802.11n~\cite{80211nstd} uses packet aggregation, whereby
multiple packets destined to the \emph{same} receiver are transmitted together within a single large
frame. In this way, the overheads associated with a single transmission are amortised
across multiple packets and higher throughput efficiency is achieved, \emph{e.g.} see \cite{afr}.

A logical extension is to consider aggregation of packets destined
to \emph{different} receivers into a single large frame.  Such
multi-destination aggregation is currently the subject of much
interest because with the increasing number of WiFi hotspots and other accessing technologies available, for a single WLAN AP, there simply may not be enough traffic to an individual destination to allow large packets to be formed in a timely manner and so efficiency gains to be realised.
One of the key issues in multi-destination aggregation is the
choice of {\color{black}{M}}odulation and  {\color{black}{C}}oding  {\color{black}{S}}cheme (MCS) for aggregated packets. Although
multi-destination aggregation allows simultaneous transmission to
multiple receivers, the channel quality between the transmitter
and each receiver is generally different, and thus the optimal MCS which matches the channel quality of each
receiver is also different. The current 802.11 standard constrains
transmitters to use the same MCS for all bits within a frame, and the state of the art is to send
multicast/broadcast packets (which contain messages for multiple
receivers) at the highest MCS rate which the receiver with the
worst channel quality can support~\cite{80211std2007}.  While this ensures that every receiver is
capable of decoding the received packet, clearly it is highly inefficient.

In this paper we consider an alternative approach to multi-destination aggregation, which still uses the same MCS for every symbol within an aggregated frame (and so does not require hardware changes) but
encodes packets destined to different receivers with different
levels of protection by using higher-layer coding
techniques.
\begin{figure}
\centering
\includegraphics[width=0.55\columnwidth, height=4cm]{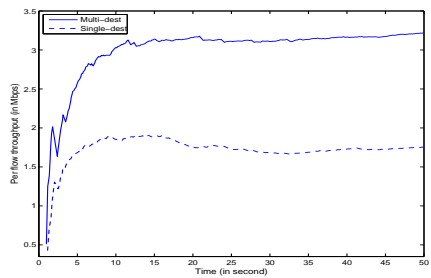}
\caption{Access point (AP) downlink throughput with single
destination aggregation and with multi-destination aggregation.
The AP is sending traffic to 10 client stations, one flow per
station. Meanwhile each station has a competing uplink flow to the
AP. Each flow has Poisson distributed packet arrivals at rate
2000. 802.11g WLAN (see Table \ref{parameters} for PHY/MAC
parameter values), AP buffer size is 200 packets, $ns2$
simulation.}\label{fig:introexample}
\end{figure}
The approach builds on an experimental observation that packets discarded at 802.11 MAC layer due to CRC errors actually contain a high proportion of correct bits, and thus potentially provide a useful channel through which
information can be transmitted.  Recently~\cite{80211Hybrid} indicates that this channel can be accurately modeled as a binary
symmetric channel. Based on this, multi-destination aggregated packets from the AP form a binary
symmetric broadcast channel between the transmitter and multiple
receivers.  Then by using appropriate BSBC-based error correction
coding bits within a single frame can be transmitted to
different destinations at different information rates while still
using the same MCS.  To our knowledge, we present the first
detailed analysis of multi-user coding for aggregation in 802.11 WLANs.  

We demonstrate in Section~\ref{theoreticalperformance} and
Section~\ref{NS2performance} that by using this coding approach
for multi-destination aggregation increases in network throughput
of more than 100$\%$ are possible over a wide range of channel
conditions.    This is illustrated, for example, in
Fig.~\ref{fig:introexample} which presents throughput measurements
for downlink transmissions in a WLAN containing 10 downlink flows
and 10 competing uplink flows.    When single destination
aggregation is used, on average insufficient packets are available
for each destination to allow a full sized frame (65535 bytes) to
be assembled. On average only 36 packets are assembled in each
single destination aggregated frame, resulting in a substantial
loss of network efficiency. At each transmission opportunity, the
AP first checks the destination address of the first packet in the
queue, and then searches through the queue to assemble packets
destined to the same receiver. With multi-destination aggregation,
full-sized frames can be assembled at every transmission
opportunity. On average 117 packets are assembled in each
multi-destination aggregated frame. Since the coding proposed here
is introduced above the MAC layer, there is no need for any
hardware changes and these performance gains therefore essentially
comes for ``free''.


{\color{black}
\section{Related work}

The concept of Multiple Receiver Aggregate (MRA) was first proposed by the TGnSync group in~\cite{TGnSync}. The idea of aggregating multiple packets into a single large frame, and then multicasting/broadcasting it to distinct receivers became the subject of much interest soon for delay-sensitive and short-packet applications such as VoIP~\cite{Lee2008BoostingVC},~\cite{VoIP2},~\cite{Softspeak},~\cite{voip3}. For example, ~\cite{VoIP2} proposes a voice multiplex$-$multicast (M$-$M) scheme of multiplexing packets from several
VoIP streams into one multicast packet for downlink transmissions to overcome the heavy overhead of VoIP traffic over WLANs. Similarly~\cite{Lee2008BoostingVC} proposes a congestion-triggered downlink aggregation scheme by stretching the 802.11n A-MPDU format~\cite{80211nstd} to carry MPDUs addressed to different destinations. Aggregation is performed only when there is congestion. When an aggregation is triggered, the VoIP packets queued at MAC layer are put into the aggregated frame in the same order as in the queue, with no sorting and no packaging for per destination. The aggregation complexity and overhead is thus reduced compared to the per-destination grouping strategy as proposed in~\cite{TGnSync} .  Apart from the downlink multi-user aggregation,~\cite{Softspeak} presents a complimentary uplink
aggregation technique that effectively serializes channel
access in the uplink direction. The combination of 
uplink and downlink aggregation mechanisms simultaneously improves
VoIP call quality while preserving network capacity
for best-effort data transfer.

All of the above works only consider homogeneous networks, i.e. stations in a WLAN have the same channel qualities and thus use the same data rate.  In a heterogeneous network where stations have different optimal transmission rates, multicasting or broadcasting the entire aggregated frame at the low enough rate to ensure all the stations can receive it will result in a significant loss in throughput. This problem is addressed in~\cite{MultiRateAgg}. This paper proposes a scheme called Data Rate based Aggregation (DRA) which groups packets in the MAC queue in terms of data rates, and then aggregates packets
for all links that have the same data rate and allows packet reordering. Such a way mitigates the performance demotion caused by aggregating across data rates. But the grouping strategy does not always provide the best performance. ~\cite{MultiRateAgg} also proposes a scheme Data Rate based Aggregation with Selective Demotion (DRA-SD) which allows a cross rate merge of two DRA frames under some conditions. The simulation results show evidence of better performance in terms of transmission time. 

Packet aggregation is considered together with network coding in~\cite{LAPNC}. This paper proposes a scheme that uses length aware packet aggregation
and network coding to improve the throughput of single relay
multi-user wireless networks. At the relay node, upload and download packets are exclusive \emph{OR}ed and then broadcast to the next hop. Aggregation is performed before coding if packets in both directions do not have similar sizes. The network coding is a packet-level coding scheme.

To the best of our knowledge, this is the first work that uses bit-level coding schemes to solve the problem of multi-rate throughput compromise in multi-destination aggregation. The proposed method could benefit from both aggregation and bit-level channel capacity improvement.

}

\section{Preliminaries}

\subsection{Multi-destination aggregated frames form a binary symmetric broadcast channel}

\begin{figure}
\centering
\includegraphics[width=0.5\columnwidth,height=2.5cm]{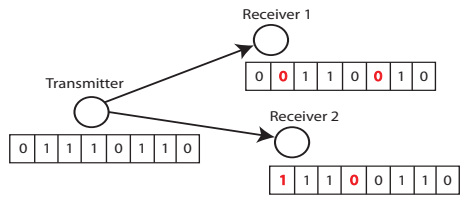}
\caption{Illustrating binary symmetric broadcast channel.   A
binary vector broadcast by the transmitter is overheard by two
stations. Reception may be lossy, with bits being received flipped
at receiver $i$ with probability $p_i$ (example bit flips are
indicated in red bold).
 }\label{fig:schematic}
\end{figure}

In a Binary Symmetric Channel (BSC) each received packet is
considered as a binary vector in which an unknown subset of bits
have been independently ``flipped'' with crossover probability
$p$.    It is shown in \cite{80211Hybrid} that, after some pre-
and post-processing, this  accurately models the behavior of the
channel provided by 802.11 corrupted frames.    In a Binary
Symmetric Broadcast Channel~ (BSBC)~\cite{Cover}, $n$ receivers overhear a
transmission.  Each receiver obtains a separate copy of the
transmission, with received bits being flipped independently with
probability $p_i$ at receiver $i$.   The crossover probability
$p_i$ embodies the link quality between the transmitter and
receiver $i$, and in general is different for each receiver and
varies with the MCS used for the transmission.  This is
illustrated schematically in Figure \ref{fig:schematic}.

\subsection{Running example: two-class WLAN}\label{twoclass}
We will use the following setup as a running example.  Namely, an 802.11
 WLAN with an AP and two classes of client stations,
$n_1$ stations in class 1 and $n_2$ in class 2.    Stations in class 1
are located relatively far from the AP and so have lossy reception with crossover probability $p$ which depends on the MCS used.   Stations in class 2 are located close to the AP and experience loss-free reception (the crossover probability is zero) for every available MCS.    Our analysis can, of course, be readily generalised to encompass situations where each station has a
different crossover probability, but the two-class case is sufficient to capture the
performance features of heterogeneous link qualities in a WLAN.

\subsection{Coding for binary symmetric broadcast channels}
The binary symmetric broadcast paradigm allows transmission of a
multi-destination aggregated frame at different information rates
to different destinations while using a single MCS.  We consider
two main approaches for achieving this, namely superposition
coding and time-sharing coding.

\subsubsection{Superposition coding}
Superposition coding works as follows.   Encoding is straightforward: binary vectors destined to different receivers are simply added together, modulo 2, and transmitted as a single binary vector.   Receiver $i$ then receives its binary vector with bits flipped by (i) the physical channel and (ii) by the addition of the messages for other receivers.    Let $p_i$ denote the physical channel crossover probability at receiver $i$ and $q_j$, $j\ne i$ denote the effective crossover probability induced by adding the message intended to receiver $j$.
Letting $r_i$ denote the probability that a bit is flipped, the  channel capacity to receiver $i$ is then
    $C_i=1-H(r_i)$, where  $H(r_i)=-r_ilog_2r_i-(1-r_i)log_2(1-r_i)$ is the binary entropy function.

For example, with $n=2$ receivers, $r_1=q_2(1-p_1)+(1-q_2)p_1$ and
$r_2=q_1(1-p_2)+(1-q_1)p_2$.  Provided messages to receiver $i$
are sent at less than this information rate, they can be
successfully decoded.      Specifically, this rate can be achieved
using the following nested decoding procedure:   (i) order the $n$
receivers by increasing crossover probability (decreasing channel
quality), with ties randomly broken, (ii) set $i=1$,
decode\footnote{ Any coding approach for a binary symmetric
channel can be used to encode the messages to receiver $i$ here
\emph{e.g.} capacity achieving codes for BSBCs are described
in:~\cite{Hsu2008},~\cite{Arikan2009},~\cite{Hari2012}.} the message for
receiver $i$ and subtract it from the received binary vector and
(iii) set $i\leftarrow i+1$ and repeat until $i$ equals the index
of the current receiver.

Although the capacity of general binary broadcast channels remains unknown, for many important special cases (e.g. for stochastically degraded binary broadcast channels), it is known that superposition coding is capacity-achieving~\cite{superposition}.

With superposition coding, the achievable sum-capacity of a binary
symmetric broadcast channel with $n$ receivers is  $\sum_{i=1}^n
C_i$ with $C_i=1-H(r_i)$ and $r_i$ the effective cross-over
probability of the binary symmetric broadcast channel between the
transmitter and receiver $i$.  For our running example of a WLAN
with two classes of stations, with $n_1=1=n_2$ the effective
cross-over probability $r_1$ for the class 1 station is
$r_1=\beta(1-p(R))+(1-\beta)p(R)$ where $p(R)$ is the crossover
probability of the physical binary symmetric broadcast channel between the
transmitter and the class 1 station (which depends, of course, on
the MCS rate $R$ selected), and $\beta$ is the crossover
probability determined by the binary addition with the message
destined to class 2. $H(\beta)$ is the information rate at which
data is transmitted to the class 2 station.

\subsubsection{Time-sharing coding}

From the discussion above it can be seen that superposition
decoding can be a relatively complex operation. A simpler but
demonstrably near-optimal choice is time-sharing
coding~\cite{Cover}.    In time-sharing, the transmitted binary
vector is partitioned into $n$ subsets of bits, where $n$ is the
number of receivers, and the $i$'th subset of bits contains the
message intended for receiver $i$ and this message is encoded at a
rate which is matched to the channel between the transmitter and
receiver $i$.   This approach is akin to packet aggregation, but
with each packet carrying a payload that is separately encoded by
the application layer$^1$.  The application layer encoding adds
appropriate redundancy that allows the intended receiver to decode
the embedded information message even when the packet is received
with bits flipped.   For the two-class WLAN example, in
time-sharing coding each transmitted frame is partitioned into two
parts,  the first intended for class 1 stations and the second
intended for class 2 stations. The portion intended for class 2
will be received
 error-free and thus does not need further protection.
The portion intended for class 1 is protected by a suitable BSBC error correcting code that
allows information to be extracted even when some bits
are corrupted; the information rate is obviously reduced compared
to a noise-free channel.

\section{Unicast throughput modelling}\label{unicast}

In this section we develop a detailed theoretical throughput
performance analysis for three multi-destination aggregation
approaches: i) uncoded frame aggregation in a packet erasure
channel paradigm; ii) aggregation with superposition coding in a
broadcast BSBC paradigm; iii) aggregation with
time-sharing coding in a  broadcast BSBC
paradigm.  We focus on the two-class setup introduced in Section
\ref{twoclass}, the extension to more than two classes being
straightforward.

\subsection{802.11 MAC model}
We consider a WLAN consisting of an access point (AP), $n_1$ class
1 stations and $n_2$ class 2 stations. We assume that all stations
are saturated (unsaturated operation is considered later). The AP
transmits ${n}_1+{n}_2$ downlink unicast flows. Namely, one flow
destined to each of the $n_1$ class 1 stations and one flow
destined to each of the $n_2$ class 2 stations. When transmitting,
the AP aggregates these downlink flows into a single large MAC
frame which is sent at a single PHY rate.  Each client station
also transmits an uplink flow to the AP.

Following Bianchi~\cite{Bianchi},~\cite{David_TON_2007}, time is divided
into MAC slots (which can be idle, success or collision slots).
Let $\tau_0$ denote the probability that the AP attempts a
transmission in a MAC slot, $\tau_1$ the probability that a class
1 station attempts a transmission and $\tau_2$ the probability
that a class 2 station attempts a transmission.    Transmissions
by class 1 stations are subject to collisions with transmissions
by the other stations in the WLAN and, in the packet erasure
paradigm, are also subject to noise-related erasures.  The
probability that a transmission from a class 1 station fails (due
to collision and/or loss) is
\begin{equation}\label{16}
    {p_f}_1=1-(1-{p_c}_1^{U})(1-{p_e}_1^{U})
\end{equation}
where ${p_e}_1^{U}$ is the probability that an uplink transmission by a
class 1 station is erased due to noise, and ${p_c}_{1}^U$ is the
probability that it collides with another transmission, with
\begin{equation}\label{17}
    {p_c}_1^{U}=1-(1-\tau_1)^{n_1-1}(1-\tau_2)^{n_2}(1-\tau_0)
\end{equation}
Class 2 stations do not suffer from noise caused erasures.
Although sub-frames destined to class 1 in the packet erasure
paradigm are subject to noise-related erasures, sub-frames
destined to class 2 are error-free. The transmission failures from the AP or a class 2 station are only caused by collisions. Hence class 2 and the AP share the same station attempt probability, i.e. $\tau_0=\tau_2$, and the same probability that a
transmission fails, which is 
\begin{equation}\label{18}
{p_f}_0={p_f}_2=1-(1-\tau_1)^{n_1}(1-\tau_2)^{n_2}
\end{equation}

The usual Bianchi~\cite{Bianchi} expression gives a relation
between the station transmission attempt probability $\tau$ and the probability $p_f$
that a transmission fails.  However, here we make use of
expression~(\ref{18new}) that extends the Bianchi expression
to take account of a finite number of retransmission attempts and
losses due to decoding errors~\cite{tianji}.
\begin{equation}\label{18new}
\tau=
\begin{cases}
   \frac{2(1-2p_f)(1-p_f^{m+1})}{M}  & m \leq m',
   \\
   \frac{2(1-2p_f)(1-p_f^{m+1})}{M+W2^{m'}p_f^{m'+1}(1-2p_f)(1-p_f^{m-m'})}  & m > m'.
\end{cases}
\end{equation}
in which $M=(1-p_f)W(1-(2p_f)^{m+1})+(1-2p_f)(1-p_f^{m+1})$,
$W=CW_{min}$, $m$ denotes the 802.11 retry limit number, and $m'$
represents the number of doubling the CW size from $CW_{min}$ to
$CW_{max}$.

\subsection{Network throuhgput} The network throughput is
\begin{equation}\label{40}
     S={\frac{X_0+{n}_1X_1+{n}_2X_2}{{E_T}}}
\end{equation}
where $X_0=\tau_0(1-{p_f}_0)(E_1^D+E_2^D)$,
$X_1=\tau_1(1-{p_f}_1)E_1^U$, $X_2=\tau_2(1-{p_f}_2)E_2^U$ with
$E_1^D$ the expected payload delivered from the AP to class 1
stations and $E_2^D$ to class 2 stations, $E_1^U$ the expected
payload delivered from a class 1 station to the AP, $E_2^U$ the
expected payload delivered from a class 2 station to the AP, and
$E_T$ is the expected MAC slot duration.  It is important to
stress that the expected payload delivered need not equal to the raw
frame payload due to the impact of corruption of the frame payload
during transmission across the radio channel and due to the
overhead of any error-correction coding.   Calculations of the
expected payloads delivered and of the expected MAC slot duration
are discussed in detail below for each of the three
multi-destination aggregation schemes considered.

\subsection{Fairness}

Before proceeding to the calculation of the flow throughputs for
the three multi-destination aggregation approaches, we note that
to ensure a fair comparison amongst different schemes it is not
sufficient to simply compare the sum-throughput. Rather we also
need to ensure that schemes provide comparable throughput
fairness, as an approach may achieve throughput gains at the
cost of increased unfairness. In the following we take a max-min
fair approach and impose the fairness constraint that all flows
achieve the same throughput. Extension of the analysis to other
fairness criteria is, of course, possible.

\subsection{Expected payload}

We begin by calculating the expected payload in a MAC
slot for the three multi-destination aggregation approaches.

\subsubsection{Uncoded} Similarly to the approach used in 802.11n
A-MPDUs~\cite{80211nstd}, we consider a situation where messages
addressed to distinct destinations are aggregated together to form
a single large MAC frame.  We do not present results here without
aggregation since the throughputs are strictly lower than when
aggregation is used~\cite{afr}.

We need to calculate the expected delivered payloads $E_1^{D}$,
$E_2^{D}$, $E_1^{U}$ and $E_2^{U}$.

We proceed as follows.  The expected payload delivered by an
uplink packet of a class 1 station is
\begin{equation}\label{21}
    E_1^U=x_1^U(1-{p_u}(R))^{L_1^U(R)}
\end{equation}
where ${p_u}(R)$ is the first-event error probability of Viterbi decoding ~\cite{Viterbi} for convolutional codes used in 802.11 standards
when transmissions are made at PHY rate $R$,
\begin{equation}\label{20}
{L_1^U(R)}={DBPS(R)}
\big\lceil\frac{({x_1^U}+L_{machdr}+L_{FCS})\times8+22}{DBPS(R)}\big\rceil
\end{equation}
is the class 1 uplink frame size in bits. $DBPS(R)$ represents
data bits per symbol at PHY rate $R$. $L_{machdr}$ is the MAC
header in bytes, and $L_{FCS}$ is the FCS field size in bytes.
$x_1^U$ is the class 1 uplink frame payload in bytes. As
transmissions by class 2 stations are erasure-free at all supported
PHY rates, the expected payload of an uplink packet from a class 2
station is
\begin{equation}\label{22}
  E_2^U=x_2^U
\end{equation}
where $x_2^U$ is the payload size in bytes of class 2
transmissions.  Turning now to the AP, similar to the approach
used in 802.11n, the aggregated MAC frame consists of
${n}_1+{n}_2$ unicast packets.   The length of a MAC frame is
\begin{equation}\label{L}
L=n_1 L_1^D + n_2 L_2^D
\end{equation}
in which $L_1^D=x_1^D+L_{subhdr}+L_{FCS}$ and
$L_2^D=x_2^D+L_{subhdr}+L_{FCS}$ are respectively the sub-frame
size for class 1 and class 2 in bytes. $L_{subhdr}$ is the
sub-header length. $x_1^D$, $x_2^D$ denote, respectively, the AP
payload size in bytes destined to class 1 and class 2 stations.
Note that the downlink PHY rate is determined by the client which
has the worst link quality, and so equals the class 1 PHY rate
$R$.  The expected payload delivered to a class 1 station by an AP
frame packet is therefore
\begin{equation}\label{24}
E_1^{D}=x_1^D(1-p_u(R))^{8\cdot L_1^D}
\end{equation}
while the expected payload delivered to a class 2 station is
\begin{equation}\label{25}
    E_2^D=x_2^D
\end{equation}

For max-min fairness we need to equalize the throughput of each
flow.  That is,  we require
\begin{align}\label{26}
    x_2^U = x_2^D=x_1^D(1-{p_u}(R))^{8\cdot L_1^D}
\end{align}
\begin{equation}\label{28}
    {\tau_1}(1-{p_f}_1)(1-{p_u}(R))^{L_1^U}{x_1^U}={\tau_2}(1-{p_f}_2){x_2^U}
\end{equation}
For a given PHY rate $R$ and AP frame size $L$ we can solve
(\ref{26}) and (\ref{L}) to obtain $x_1^D$ and $x_2^D$. {\color{black}As
${p_f}_1$ depends on the payload size $x_1^U$ due to noise-related
erasures, we need to solve (\ref{28}) jointly  with the MAC model (\ref{18new})
to obtain $x_1^U$, $\tau_1$ and $\tau_2$.  We can then obtain $E_1^D$, $E_2^D$,
$E_1^U$, $E_2^U$ from (\ref{24}), (\ref{25}), (\ref{21}),
(\ref{22}) as required.}

\subsubsection{Time-sharing coding}

For the binary symmetric broadcast paradigm we start by
considering the simpler time-sharing coding scheme.  As in the
erasure channel case, MAC frames are constructed by aggregating
two portions: one intended for class 1 stations and protected by
an application layer error correction code (with coding rate
matched to the channel quality between the AP and class 1
stations),  the second intended for class 2 stations and uncoded
(since the PHY layer MCS provides adequate protection).
 Each portion is further sub-divided into packets intended for the
different stations. We also apply similar coding to protect uplink
transmissions from class 1 stations to allow information to be
recovered from corrupted uplink frames.

Let $x_1^D$ denote the downlink information payload size for a
class 1 station and $x_2^D$ for a class 2 station. Suppose a
downlink PHY rate $R$ is chosen and the crossover probability for
class 1 stations is $p(R)$.  The number of coded bytes to ensure
reception of $x_1^D$ information bytes is $x_1^D/{(1-H(p(R)))}$.
The expected downlink payload delivered to class 1 and class 2 are
$E_1^D=x_1^D$ and $E_2^D=x_2^D$. To equalize the downlink
throughputs of stations in both classes (i.e. for max-min
fairness), we therefore require
\begin{equation}\label{43}
E_1^D=E_2^D
\end{equation}
The AP frame size is $L=n_1 L_1^D + n_2 L_2^D$ where $L_1^D=
(x_1^D+L_{subhdr}+L_{FCS})/(1-H(p(R))) )$,
$L_2^D=x_2^D+L_{subhdr}+L_{FCS}$.

To equalize the uplink and downlink throughputs we require
\begin{align}
E_2^U=E_1^U=E_1^D
\end{align}
The expected uplink payload delivered from class 1 and class 2 are
$E_1^U=x_1^U$ and $E_2^U=x_2^U$. Hence we have
$x_2^U=x_1^U=x_1^D=x_2^D$ (where we are making use here of the fact
that since frames are not erased in the binary symmetric broadcast
channel paradigm, ${p_e}_1^U=0$ and thus $\tau_0=\tau_1=\tau_2$).
Therefore given a specified AP frame size $L$ we can solve for $\tau_1$ and $x_1^D$ in the similar way and
obtain $E_1^D$, $E_2^D$, $E_1^U$, $E_2^U$.

\subsubsection{Superposition coding}

With superposition coding the MAC frames are constructed in two
steps. Once a value of $\beta$ has been determined, binary vectors
are generated by aggregating IP packets of each class, and these
are then summed, modulo 2, to generate the MAC frame. Despite the
coding scheme being more complicated, the throughput analysis is
similar to the time-sharing case. The main difference lies in the
calculation of the downlink payload size.

Letting $R$ denote the downlink PHY rate used by the AP and $p(R)$
denote the corresponding BSC crossover probability.   The downlink
BSBC capacity in bits per channel use between the AP and a class 1
station is $1-H(\beta\circ{p(R)})$, where $\beta\circ{p(R)}=\beta(1-p(R))+(1-\beta)p(R)$, and that between the AP and a
class 2 station is $H(\beta)$. The AP frame payload is formed by
superimposing ${n}_2$ packets destined to class 2 stations to
${n}_1$ packets destined to class 1 stations.  Hence, the AP frame
size is $L=n_1 L_1^D = n_2 L_2^D$ where

\begin{equation}
L_1^D=\frac{x_1^D+L_{subhdr}+L_{FCS}} {(1-H(\beta\circ{p(R)})) }
\end{equation}

\begin{equation}
L_2^D=\frac{x_2^D+L_{subhdr}+L_{FCS}} {H(\beta)}
\end{equation}
$E_1^D=x_1^D$ is the expected downlink payload for a class 1
station, and $E_2^D=x_2^D$ is the expected downlink  payload for
a class 2 station.  To equalize the downlink throughputs of
stations in both classes, we require
\begin{equation}\label{48}
    E_1^D=E_2^D
\end{equation}
The ratio ${{n}_1}/{{n}_2}$ then fixes the value of $\beta$. With
the value of $\beta$ determined, given a specified AP frame size
$L$ we can solve to obtain $E_1^D$, $E_2^D$.    To equalize the
uplink and downlink throughputs we then require
\begin{equation}
E_2^U=E_1^U=E_2^D
\end{equation}

\subsection{Expected MAC slot time}
Now we calculate the expected MAC slot duration.  Let $T_{AP}$
denote the duration of a transmission by the AP, $T_{1}$ the
duration of a transmission by stations in class 1 and $T_{2}$ the
duration of a class 2 transmission.   As we have seen previously,
we cannot adopt the usual approach of assuming that these
transmissions are all of equal duration. However, we can still
make use of the ordering in frame durations $T_{AP}\ge T_1 \geq
T_2$. With this ordering, there are four possible types of MAC
slot:
\noindent 1. \emph{AP transmits}: the slot duration is $T_{AP}$ (even if other stations also transmit).  The event occurs with probability $\tau_0$.

\noindent 2. \emph{Class 1 transmits}: the slot duration is $T_1$ if the AP does not transmit and at least one class 1 station transmits.  This event occurs with probability $p_{T_{1}}=(1-(1-{\tau_1})^{n_1})(1-{\tau_0})$.

\noindent 3. \emph{Only class 2 transmits}: the slot duration is $T_2$ if only class 2 stations transmit.    This event occurs with
probability $p_{T_{2}}=(1-(1-{\tau_2})^{n_2})(1-{\tau_0})(1-{\tau_1})^{n_1}$.

\noindent 4. \emph{Idle slot}: the slot duration is the PHY slot size $\sigma$ is no station transmits.  This event occurs with
probability  $p_{Idle}=(1-{\tau_1})^{n_1}(1-{\tau_2})^{n_2}(1-{\tau_0})$.

The expected MAC slot duration is therefore
\begin{equation}\label{39}
    {E_T}={p_{Idle}}\sigma+{\tau_0}{T_{AP}}+ {p_{T_{1}}}{T_{1}}+ {p_{T_{2}}}{T_{2}}
\end{equation}

\subsection{MAC overheads}

The duration of a class 1 station transmission is
$T_{1}=T(x_1^U)+T_{oh}$ where $x_1^U$ is the payload in bytes of
a class 1 station frame, and of a class 2  station transmission is
$T_{2}=T(x_2^U)+T_{oh}$ where $x_2^U$ is the  payload in bytes
of a class 2 station frame.   The duration of an AP transmission
is $T_{AP}=T(L)+ T_{oh}+T_{phyhdr1}-T_{phyhdr}$ where $L$ is the
payload in bytes of an AP frame and $T_{phyhdr1}$ the PHY/MAC
header duration for an aggregated frame.   Here,
$T_{oh}=T_{difs}+2T_{phyhdr}+T_{sifs}+T_{ack}$ is the PHY and MAC
siganlling overhead, with $T_{phyhdr}$ the PHY header duration in
$\mu{s}$, $T_{ack}$ the transmission duration of an ACK frame in
$\mu{s}$, $T_{difs}$ a DIFS and $T_{sifs}$ a SIFS.    $T(x)=4\cdot
\lceil\frac{(x+L_{machdr}+L_{FCS})\times8+22}{DBPS(R)}\rceil$ is
the transmission duration, including MAC framing, of a payload of
$x$ bytes at PHY rate $R$.

In these calculations we assume that uplink transmissions
by client stations are immediately acknowledged by the AP (rather
than, for example, using a block ACK proposed in
802.11e~\cite{80211estd}).  Similarly, we assume that downlink
transmissions are immediately acknowledged by client stations and,
to make our analysis concrete, we adopt the approach described
in~\cite{SMACK} which uses the orthogonality of OFDM subcarriers
to allow a group of client stations to transmit feedback signals
at the same time, and thereby ACK collisions are avoided. However, we stress that
these assumptions regarding ACKing really just relate to the calculation of the MAC
overheads and our analysis could be readily modified to account
for alternative acknowledgment mechanisms.

\begin{figure}
  \begin{center}
  \includegraphics[width=0.8\columnwidth,height=2.5cm]{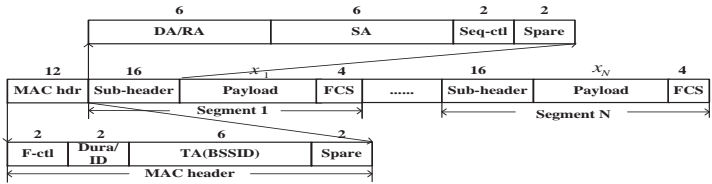}\\
   \caption{Erasure channel frame format.}\label{frameformat3}
  \end{center}
\end{figure}

Similarly, to keep our discussion concrete, we assume the frame format shown in Fig. \ref{frameformat3} is used for multi-destination aggregation in the packet erasure paradigm and with time-sharing coding. Again, it is important to stress that this just relates
to the calculation of the MAC overheads.   In
Fig. \ref{frameformat3} a sub-header is prefixed to each IP packet
to indicate its receiver address, source address and packet
sequence information. An FCS checksum is used to detect corrupted
packets in packet erasure paradigm. Since the sub-header already contains the receiver
address, source address and sequence control, the MAC header
removes these three fields, but keeps other fields unchanged from
the standard 802.11 MAC header. We assume that the
MAC header is transmitted at the same PHY rate as the PLCP header
and thus is error-free. 

{\color{black}Although the sub-header of each time-sharing segment contains the receiver address as depicted in Fig. \ref{frameformat3}, this field is not reliable due to channel noise. And as the length of each coded segment depends on the current channel quality, it varies over time. Therefore, we need to notify each receiver in the common MAC header to locate its segment. We use the spare field in the MAC header to map the initial position of each segment to its destination. Each receiver is allocated with a unique ID when associated with the AP. In the mapping field, receivers are identified using this ID number instead of their MAC addresses to save space.}

\section{Multicast throughput modelling}\label{multicast}
The foregoing unicast analysis can be readily extended to encompass multicast traffic.   The AP
now multicasts two downlink flows which are aggregated
into a single MAC frame. Flow 1 is communicated to the $n_1$ class 1
stations and flow 2 is communicated to the $n_2$ class 2 stations.  When there are no competing uplink flows we
can compute the throughput using the analysis in
Section~\ref{unicast} by selecting the following parameter values:
${n}_1={n}_2=1$; $x_1^U=x_2^U=0$; ${p_e}_1^U={p_c}_1^U=0$;
$\tau_1=\tau_2=0$; $\tau_0=2/(W+1)$. The expected payload and MAC
slot duration can now be calculated using the same method as the
unicast analysis, but for a multicast network the
per-station multicast saturation throughput is $S_1=\frac{\tau_0E_1^D}{E_T}$ for class 1 stations and $S_2\frac{\tau_0{E_2^D}}{{E_T}}$ for class 2 stations.   The network sum-throughput is $S=n_1S_1 + n_2S_2$.

\section{Theoretical Performance }\label{theoreticalperformance}

\begin{figure}
 \centering
  \includegraphics[width=0.6\columnwidth,height=4.2cm]{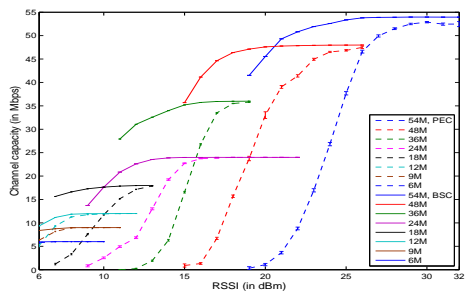}
  \caption{Experimental measurements of outdoor BSBC capacity provided by 802.11 corrupted frames~\cite{80211Hybrid}.  For comparison, also shown is the channel capacity when a packet erasure paradigm is adopted (packets failing a CRC check are discarded).}\label{outdoorchannelcapacity}
\end{figure}

We use the models developed in the previous sections to compare
the throughput performance of the uncoded and binary broadcast
schemes.    The models yield the throughput as a function of the
channel error rate, \emph{i.e.} the packet erasure rate for
uncoded operation and the bit crossover probability when using
coding. Combining these with data on channel error rate as a
function of SNR/RSSI and PHY rate allows us to determine the
optimal transmission PHY rates for downlink and uplink flows and
obtain the maximum network throughput for a range of SNR/RSSIs.
For this purpose we use the experimental channel measurements
shown in Fig.~\ref{outdoorchannelcapacity}, which are taken from
\cite{80211Hybrid}.   The experimental PEC capacity shown in
Fig.~\ref{outdoorchannelcapacity} is for a packet length of 8640
bits. To obtain the PEC capacity for any other values of packet
length, we need to first derive the first event error probability
of Viterbi decoding for convolutional codes, which is given by
\begin{equation}\label{3}
P_u=1-(1-FER)^{(1/l)}
\end{equation}
where $FER$ is the measured packet erasure rate at a given RSSI,
and $l$ is the packet length used in the experiment, \emph{i.e.}
8640 bits. Using this first event error probability $P_e$,  the
packet erasure rate for a packet length of $L$ is, in turn, given
by $1-(1-{P_u})^L$.

The MAC parameters used are detailed in Table~\ref{parameters}.

\subsection{Unicast}
\begin{figure}
\centering
\subfigure[$n_1=n_2=10$ stations]{
  \includegraphics[width=0.46\columnwidth, height=4.5cm]{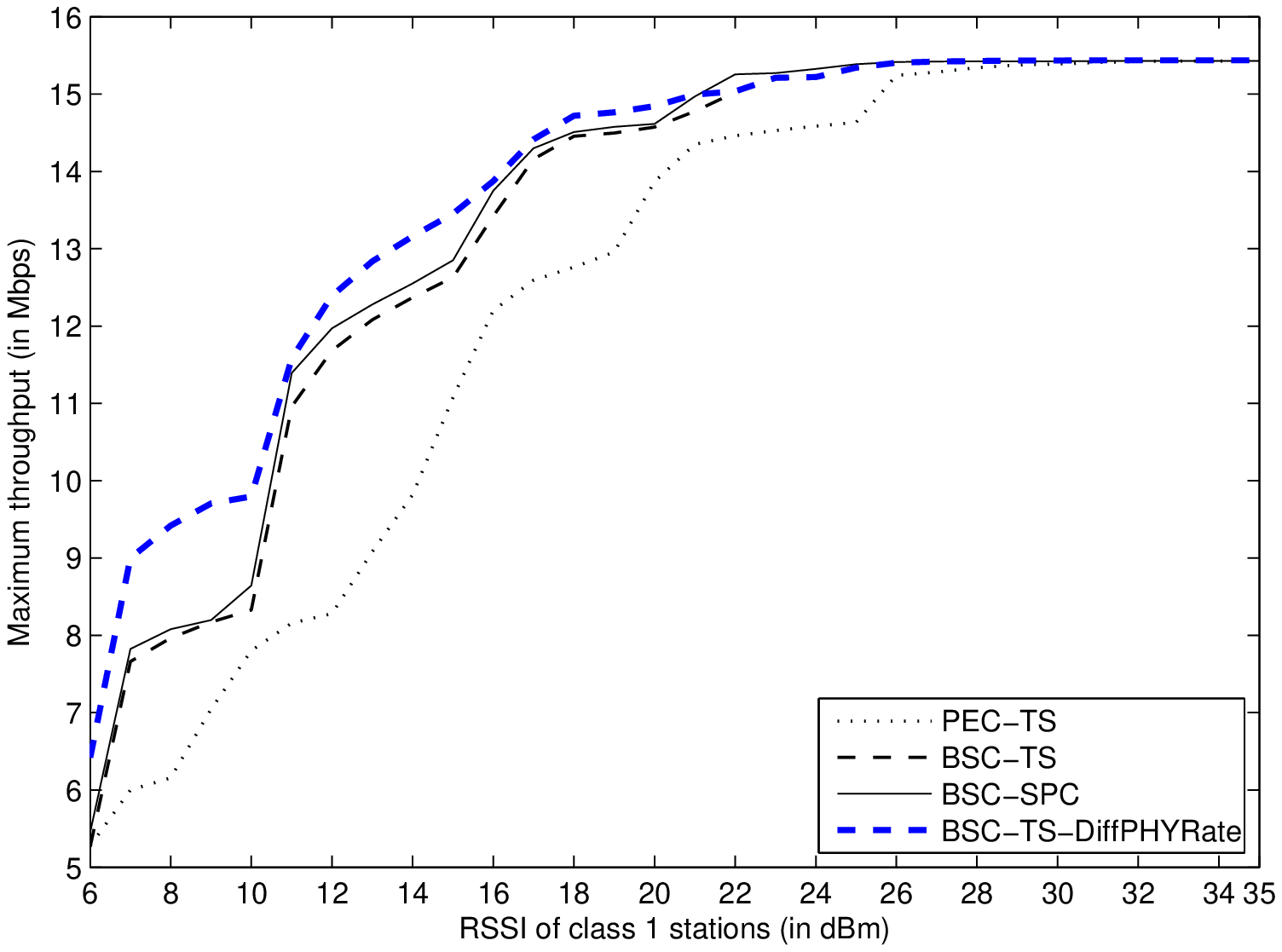}\label{unicast10}
 }
 \subfigure[$n_1=n_2=5$ stations]{
   \includegraphics[width=0.46\columnwidth,height=4.5cm]{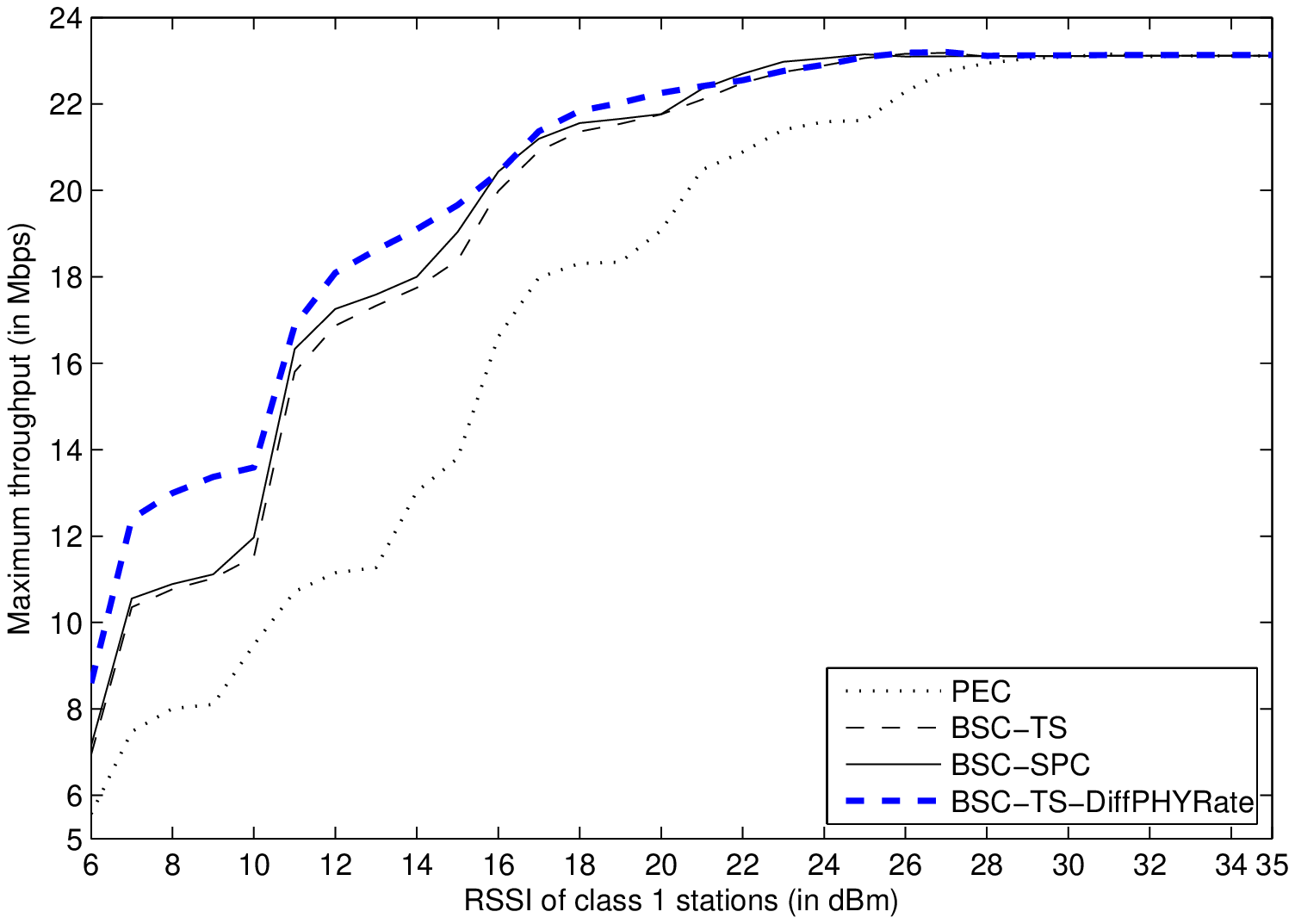}\label{unicast5}
 }
  \caption{Unicast maximum network throughput vs. RSSI of class 1 stations, $L=8000$ bytes.
   TS and SPC indicate time-sharing coding and superposition coding respectively.}
\end{figure}

We first consider unicast traffic. We compare the
throughput performance for four different approaches: 1) uncoded; 2) time-sharing coding with the entire packet
transmitted at a single PHY rate; 3)  superposition coding; 4)
 time-sharing coding with segments transmitted at different PHY
rates, \emph{i.e.} segments destined to stations in class 2 are
transmitted at the highest PHY rate available, which is 54Mbps in
802.11a/g, and the downlink PHY rate for class 1 segments is
selected to maximise the network throughput. Fig.~\ref{unicast10}
shows the sum-throughputs achieved by these different approaches
for a network consisting of 20 client stations, 10 in class 1 and
10 in class 2. This is quite a large number of saturated stations
for an 802.11 WLAN and suffers from a high level of collision
losses. Comparing it with Fig.~\ref{outdoorchannelcapacity}, it
can be seen that the throughput is significantly reduced due to
the various protocol overheads and collisions that have now been
taken into account. Nevertheless, the relative throughput gain of
the coding-based approaches compared to the uncoded approach
continues to exceed $50\%$ for a wide range of RSSIs. Time-sharing
coding achieves very similar performance to the more sophisticated
superposition coding. The approach of using different PHY rates
for different time-sharing coding segments naturally achieves
higher throughputs than using the same PHY rate. The gains are
especially high at low RSSIs. This is because when the entire
packet is transmitted using the same PHY rate, the optimal PHY
rates for the uncoded and coded schemes are usually not very
different, \emph{e.g.} it is impossible that the uncoded scheme
chooses 6Mbps but a coded scheme chooses 54Mbps. However if
segments destined to distinct receivers are allowed to use
different PHY rates, the optimal PHY rates for both schemes can be
quite different, \emph{e.g.} in our two-class example, the portion
for class 2 always uses a quite high PHY rate of 54Mbps, while the
portion for class 1 could use a very low PHY rate, especially at
low RSSIs.

Fig.~\ref{unicast5} shows the corresponding results for a smaller
number of client stations, 5 in class 1 and 5 in class 2. The
overall throughput is higher than that with 20 stations because of
the lower chance of collisions, and the gain offered by the coding
approaches is even higher \emph{i.e.} more than 75\% over a wide
range of RSSIs.
\begin{table}\tiny
\caption{MAC protocol parameters}\label{parameters}
\centering
\begin{tabular}{c c|c c|c c}
  \hline
  $T_{sifs}$ ($\mu$s)          & 16         & $L_{subhdr}$ (bytes)        & 16 & $T_{ack}$ ($\mu$s)           & 24
  \\ \hline
  $T_{phyhdr}$ ($\mu$s)        & 20         & $L_{FCS}$ (bytes)           & 4  &  $T_{difs}$ ($\mu$s)          & 34 \\  \hline
  $T_{phyhdr1}$ ($\mu$s)       & 36         & $L_{machdr}$ (bytes)        & 24  &  Retry limit            & 7
  \\  \hline
  Idle slot $\sigma$ ($\mu$s)  & 9          & $CW_{max}$                  & 1024  & $CW_{min}$    & 16 \\ \hline
\end{tabular}
\end{table}

\begin{figure}
\centering
\subfigure[$n_1=n_2$, $RSSI=13$dBm]{
  \includegraphics[width=0.46\columnwidth, height=4.5cm]{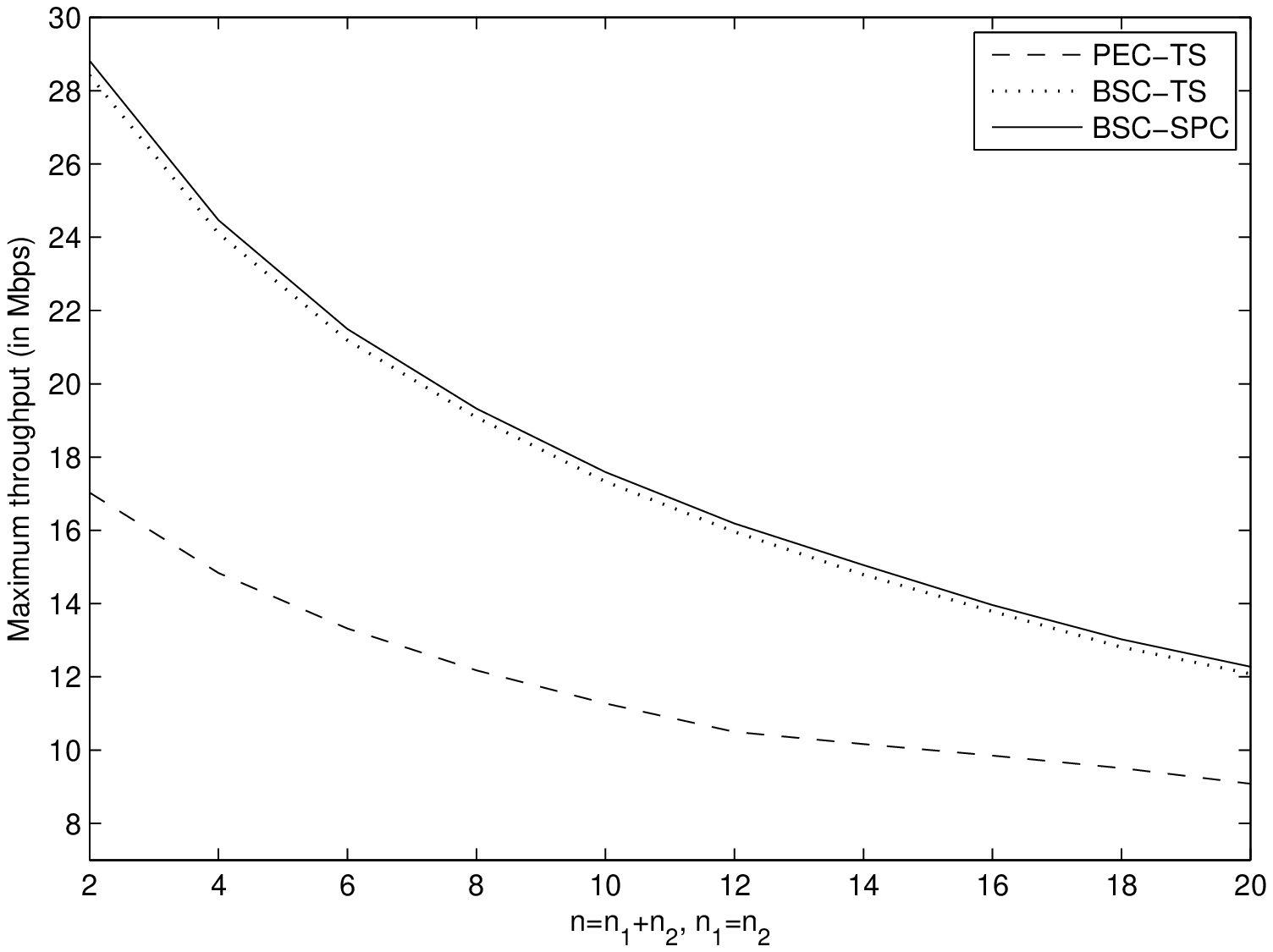}\label{unicastfixedratio}
 }
 \subfigure[$n_1+n_2=10$, $RSSI=12$dBm]{
   \includegraphics[width=0.46\columnwidth,height=4.5cm]{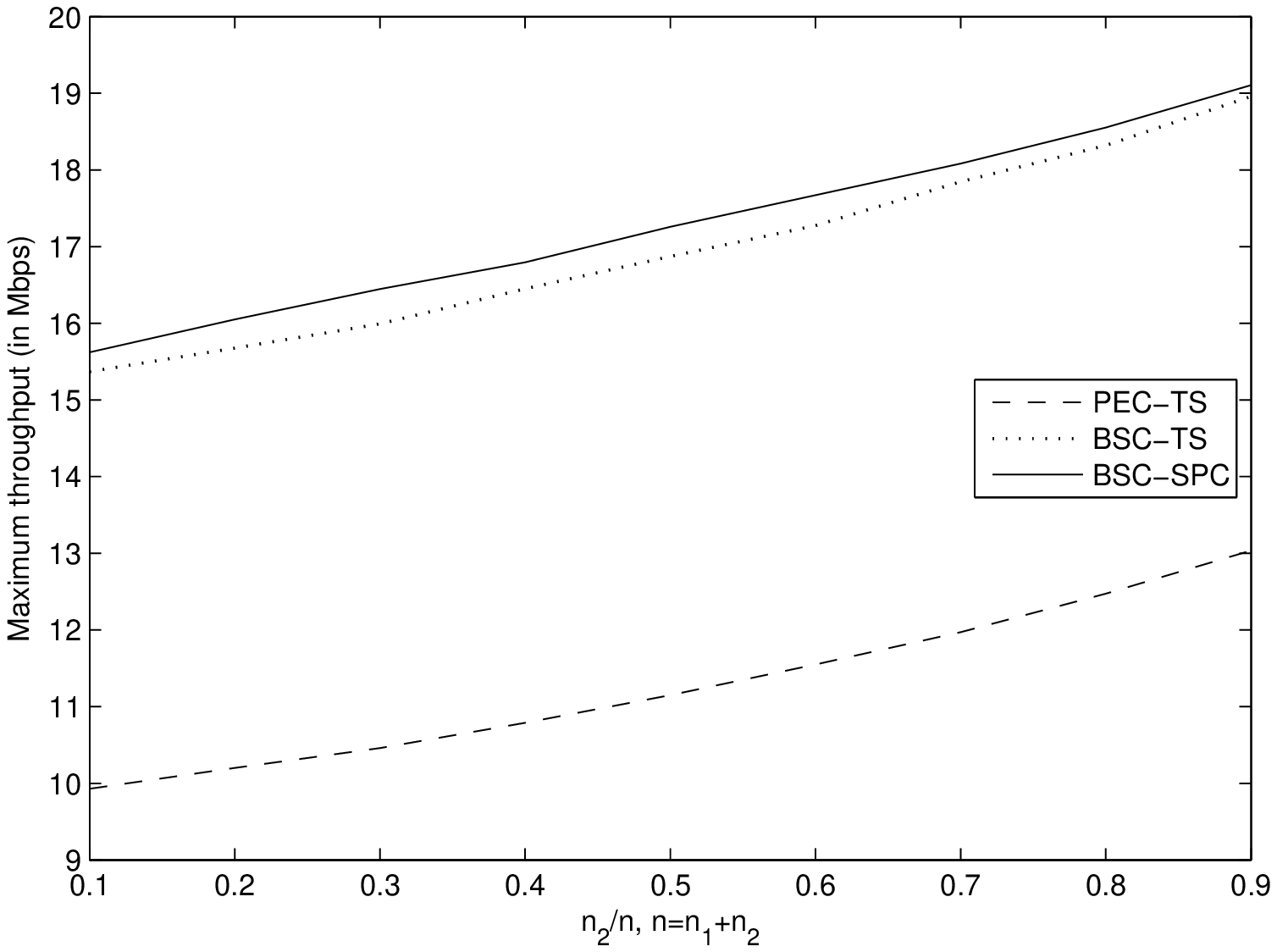}\label{unicastfixedtotal}
 }
  \caption{Unicast maximum network throughput, $L=8000$ bytes.
   TS and SPC indicate time-sharing coding and superposition coding respectively.}
\end{figure}

Fig.~\ref{unicastfixedratio} illustrates how the number of
stations affects these results. The decrease in network throughput
with increasing number of stations is evident, as is the
significant performance gain offered by the coding schemes.   For
smaller numbers of stations (which is perhaps more realistic), the
throughput gain offered by the coding approaches is larger
\emph{e.g.} nearly up to $70\%$ for 2 stations and falling to
around $30\%$ with 20 stations. The proportion of class 1 and class 2 stations can  be expected to affect the relative performance of the uncoded and coded schemes. This is because we now have multiple transmitting stations, and each station defers its contention window countdown on detecting transmissions by other stations. Since class 1 transmissions are of longer duration than class 2 transmissions, we expect that the network throughput will rise as the number of class 1 stations falls and indeed we find that this is the case. See, for example, Fig.~\ref{unicastfixedtotal} which plots the network throughput versus the varying ratio of the number of class 2 stations over the total number while maintaining the total number of client stations constant as $n_1+n_2=10$.

\subsection{Multicast}
\begin{figure}
\centering
\subfigure[$L=8000$ bytes]{
  \includegraphics[width=0.46\columnwidth, height=4.5cm]{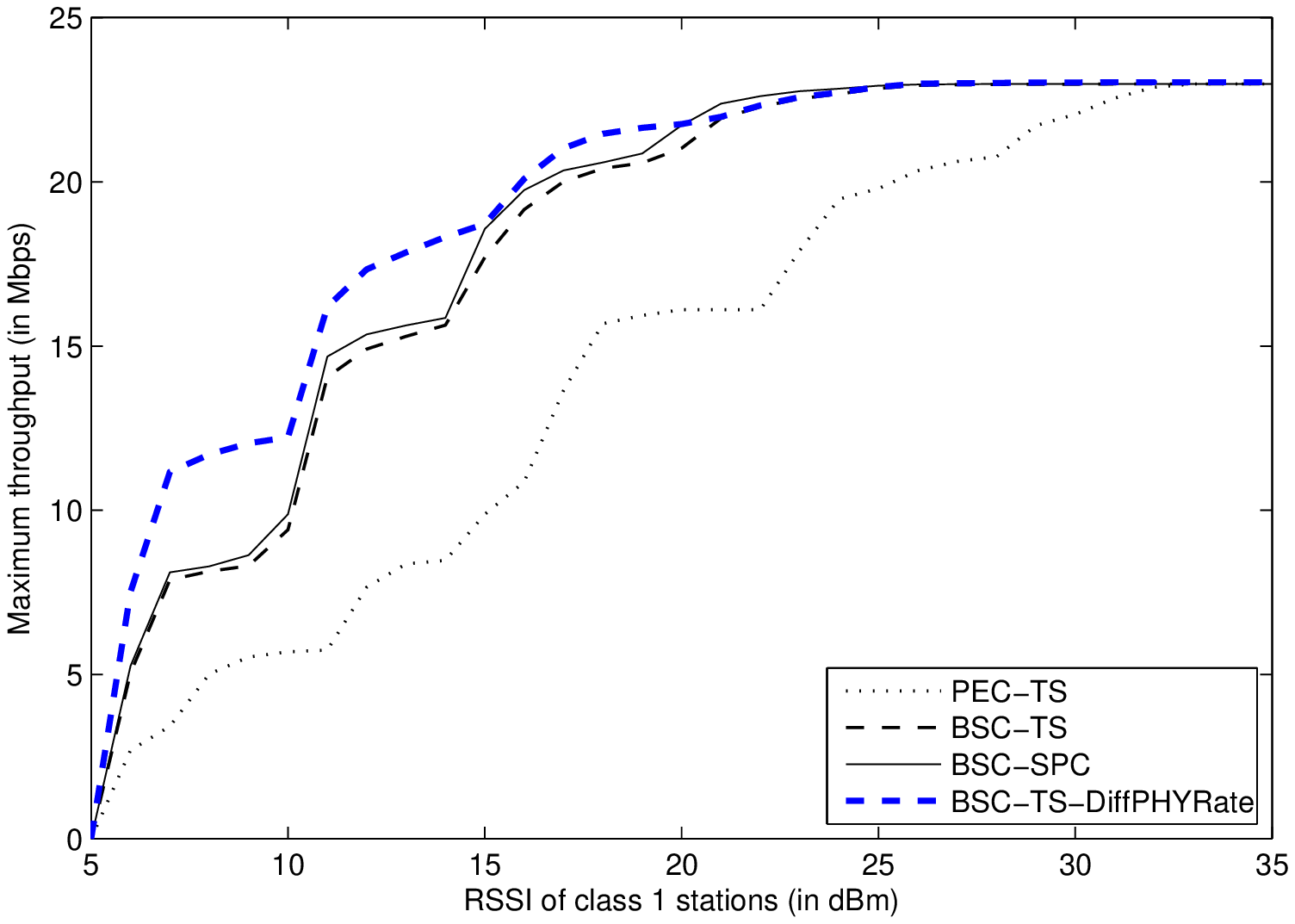}\label{multicast8000}
 }
 \subfigure[$L=65536$ byte]{
   \includegraphics[width=0.46\columnwidth,height=4.5cm]{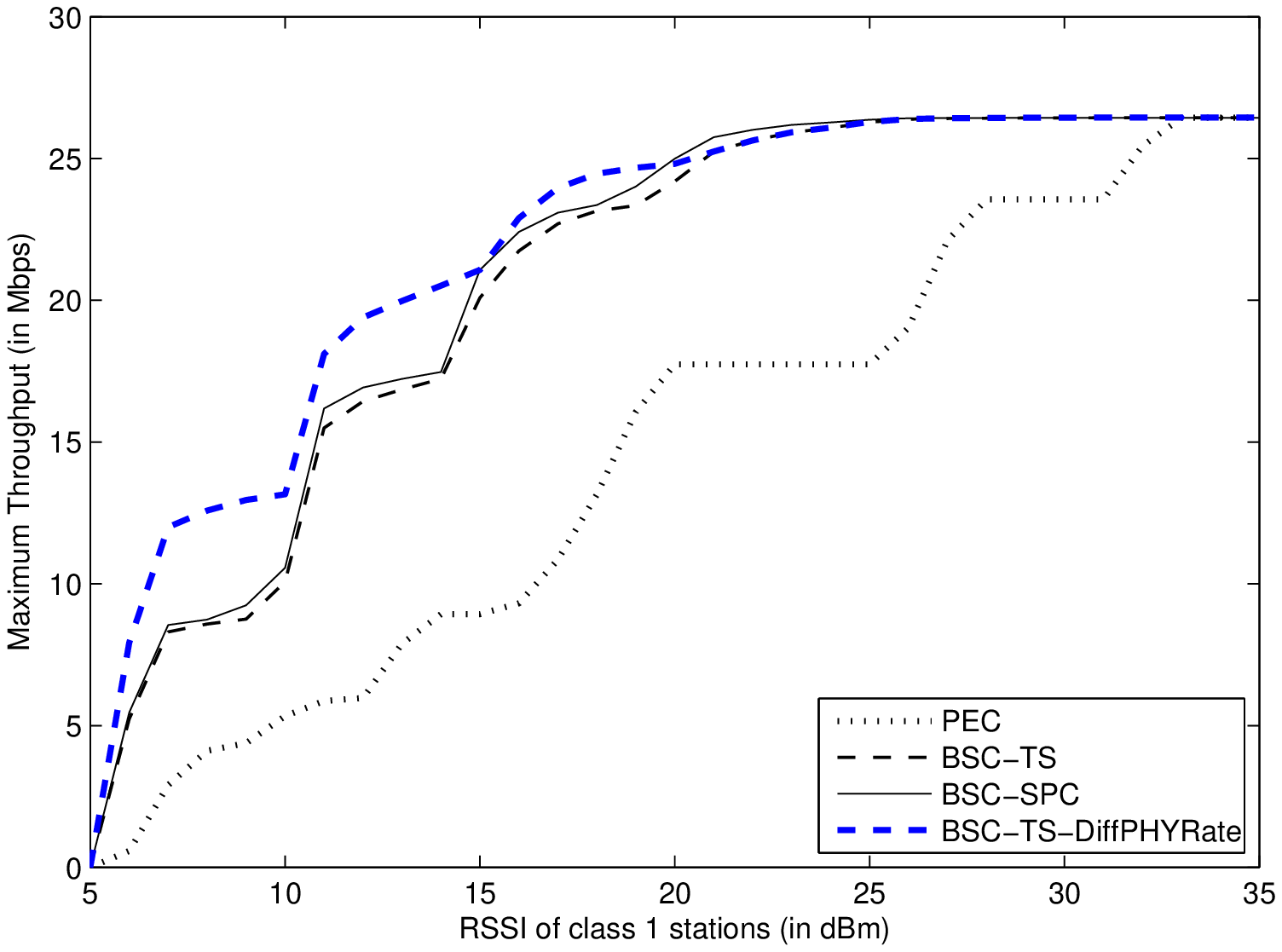}\label{multicast64K}
 }
  \caption{Multicast per-station maximum throughput vs RSSI of class 1 stations, $L=8000$ bytes, $n_1=n_2=10$ stations. TS and SPC indicate time-sharing coding and superposition coding respectively.}
\end{figure}

For multicast, we  compare the per-station throughput for the
four aggregation approaches. Fig.~\ref{multicast8000} shows the
per-station throughput for a network with $n_1=10$ class 1
stations and $n_2=10$ class 2 stations. The throughput is much
higher than the unicast case as shown in Fig.~\ref{unicast10}
because of the absence of collisions with uplink flows.
Nevertheless, both of the coded schemes (time-sharing and
superposition coding) continue to offer substantial performance
gains over the uncoded approach, increasing throughput by
almost $100\%$ over a wide range of RSSIs. The superposition
coding scheme performs slightly better than the time-sharing
scheme, but the difference is minor. Fig.~\ref{multicast64K} shows
the corresponding results with a larger MAC frame size of $65536$
bytes, which is the maximum frame size allowed in the 802.11n
standard~\cite{80211nstd}. The performance gain offered by the coded
approaches increases as the frame size is increased. Since the
per-station multicast throughput is independent of the number of
stations, we show results for only one value of $n_1$ and $n_2$.

\section{NS-2 Simulations}\label{NS2performance}
The theoretical performance results presented in
Section~\ref{theoreticalperformance} consider the scenario where stations are saturated, and
so there are always enough packets available to form  maximum-sized
aggregated packets.  It can be expected that the impact of traffic arrivals and queueing strongly affects the availability of packets  for aggregation.  In some circumstances, stations may not have enough packets to allow the maximum-sized aggregated frames to be formed (achieving the highest aggregation efficiency). In this section we use the Network Simulator 2 (NS-2) to evaluate the benefits of the proposed schemes in unsaturated situations.  It is worth noting that  the NS-2 simulations in this section are not aimed to verify the throughput models and the performance results presented  in
Section~\ref{unicast},~\ref{multicast} and \ref{theoreticalperformance}, as the theory analysis is based on the widely recognized Bianchi model which has already been thoroughly verified.  

We use two metrics to evaluate the performance:

\noindent 1. \emph{Per downlink flow throughput (in bits/s)}: Let $m_i$ denote the number of
    received packets of downlink flow $i$ during the simulation
    duration $t$. The packet length is $L$ in bytes. The throughput
    of flow $i$ is thus $8Lm_i/t$. The per downlink flow
    throughput is the mean over all $n$ downlink flows, which is $\sum_{i=1}^{n}{8Lm_i/(tn)}$.

\noindent 2. \emph{Mean downlink delay (in seconds)}: We define the delay of a packet
    as the period from when it arrives at the {\color{black}I}nter{\color{black}F}ace {\color{black}Q}ueue (IFQ) of the transmitter
    until it arrives at the MAC layer of the receiver. The mean
    downlink delay is the mean over all downlink packets. We use DropTail FIFO
    queues in our simulations.

\subsection{Single-destination vs multi-destination aggregation}
We begin by comparing uncoded multi-destination aggregation with
single-destination aggregation. We consider a WLAN
with an AP and $N$ stations. The AP has $n$
downlink unicast flows individually destined to each of the $N$
stations, and meanwhile each station has a competing uplink flow
destined to the AP. Different from the two-class example described
in Section~\ref{twoclass}, as we would like to emphasize the impact of packet availability to the two aggregation schemes, in this example we assume that all
links are error-free. Downlink transmissions are large aggregated
packets and uplink transmissions are normal 802.11 packets. As
aggregated packets are quite long, we use the RTS/CTS exchange
before data packets in our simulations. Again, we assume that the
multi-destination aggregation uses the SMACK~\cite{SMACK} scheme
to allow receivers to send acknowledgments simultaneously, and
hence there is only one ACK packet duration after each
aggregated data packet. To ensure a fair comparison, for the
single-destination aggregation, we assume that the receiver sends
one ACK after each aggregated packet to acknowledge reception of
data packets aggregated in that packet.  The traffic is real-time
stream data and follows a Poisson process with mean arrival rate of
$\lambda$. The RTP/UDP/IP header is 40 bytes (IP=20 bytes; UDP=12
bytes; RTP=8 bytes). The maximum aggregated frame size is 65535
bytes. The PHY data rate is 54Mbps.

\begin{figure}
  \begin{center}
  \includegraphics[height=6cm, width=0.75\columnwidth]{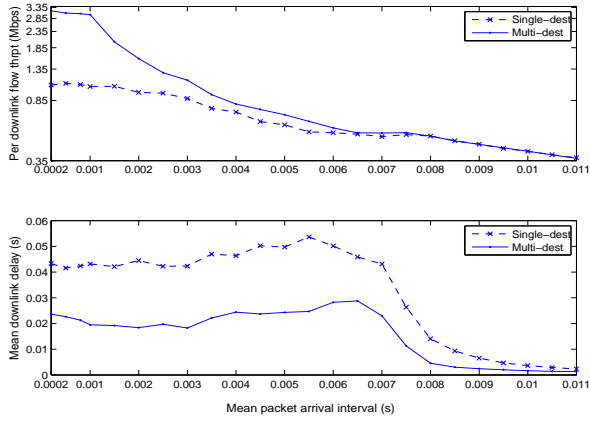}\\
   \caption{Per downlink flow throughput and mean downlink delay vs. mean Poisson packet arrival interval, 10 stations, flow packet size 500B, queue size 100 packets. }\label{10Q100P500B}
  \end{center}
\end{figure}

Fig.~\ref{10Q100P500B} plots the per downlink flow throughput and
the mean downlink delay versus the mean packet arrival interval
($1/\lambda$) for a WLAN with 10 stations. The traffic packet size
is 500 bytes. The queue size is 100 packets. It can be seen that
as expected the multi-destination aggregation achieves strictly
higher throughput and lower delay than single-destination
aggregation.   When the mean packet inter-arrival time is above
0.008s (corresponding to a light traffic load), both schemes
achieve similar throughput. This is because in this range the AP
is unsaturated, \emph{i.e.} there is typically only one packet
available to be aggregated. As the mean inter-arrival time is
decreased (and so the traffic load increases), the AP queue starts
to build up. The measured delay then includes the period awaiting
in the queue. The multi-destination aggregation scheme tends to
aggregate more packets in each transmission, and hence obtains
higher throughputs and lower delays. When the mean inter-arrival
time is decreased to 0.001s, the AP is saturated (the queue is
persistently backlogged) and it can be seen that multi-destination
aggregation achieves about a 200\% increase in throughput over
single-destination aggregation.

\begin{figure}
\centering
\subfigure[Flow packet size 500B]{
  \includegraphics[width=0.46\columnwidth, height=4.7cm]{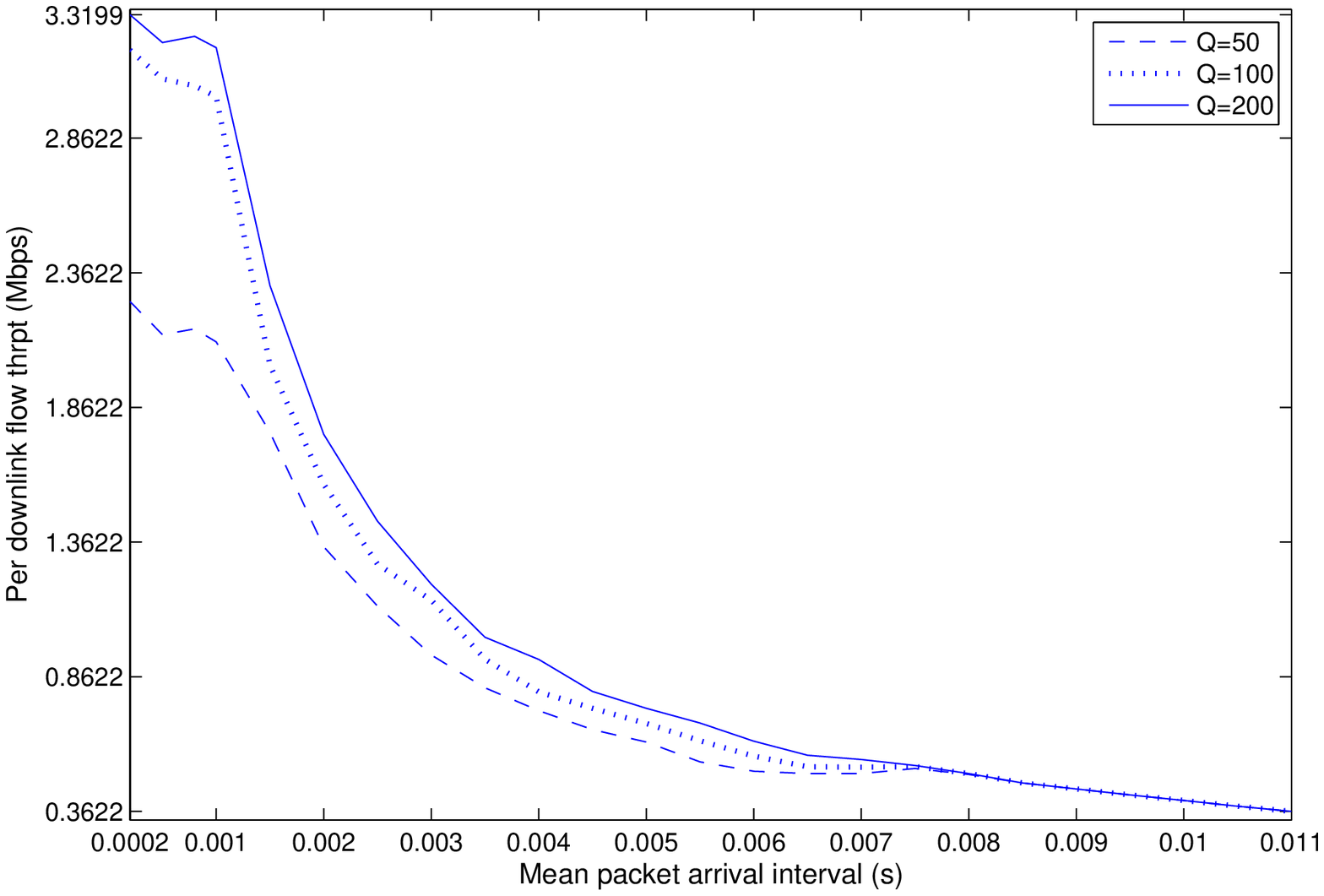}\label{Qsize}
 }
 \subfigure[Queue size 200 packets]{
   \includegraphics[width=0.46\columnwidth,height=4.3cm]{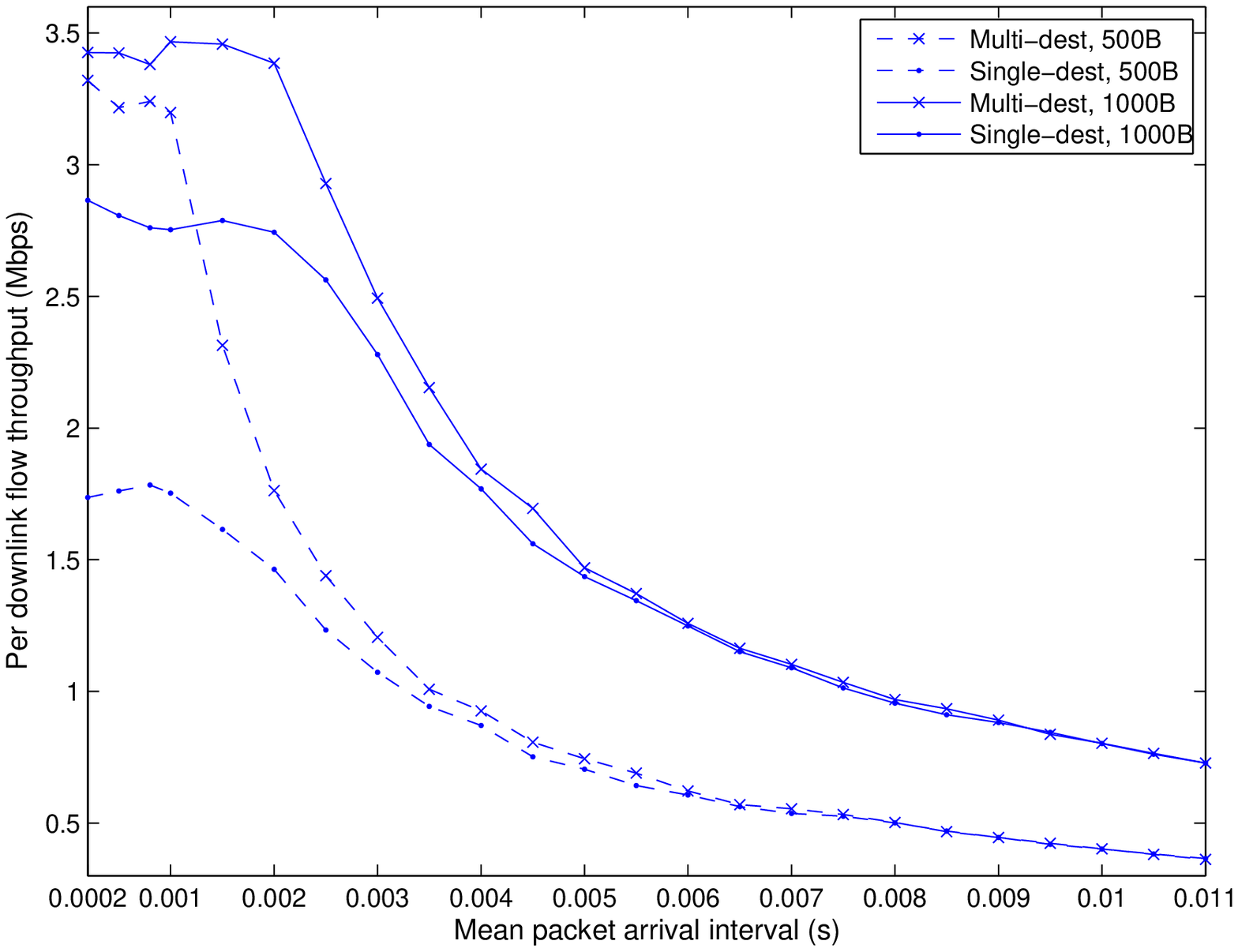}\label{Pktsize}
 }
  \caption{Per downlink flow throughput vs. mean Poisson packet arrival interval for (a) different queue sizes, (b) different packet sizes,  10 stations}
\end{figure}

Fig.~\ref{Qsize} illustrates how the queue size affects the
multi-destination aggregation throughput. Again there are 10
stations and the traffic packet size is 500 bytes. We compare
three queue sizes of 50, 100 and 200 packets. It can be seen that
when the AP becomes saturated, a larger queue provides a higher
throughput because there are more packets available to be
aggregated. When the queue of 200 packets is filled up, the
aggregated packet reaches the maximum size limit and thus less
than 200 packets are aggregated in one transmission.

Fig.~\ref{Pktsize} plots the downlink flow throughput versus the
mean packet inter-arrival time for two packet sizes of 500 bytes and
1000 bytes. There are 10 stations and the queue size is 200
packets. With single-destination aggregation, the throughput
with a packet size of 1000 bytes is around twice that with a
packet size of 500 bytes. This is because with the fixed queue
size and packet arrival rate, the expected number of packets
available to be aggregated is also fixed. However, with
multi-destination aggregation, when the queue fills, both packet sizes
obtain almost the same throughput because both reach the
maximum aggregated frame size limit.

\subsection{Uncoded vs coded approaches}

In this section we compare uncoded multi-destination aggregation with the binary broadcast
time-sharing coding scheme.  We consider the two-class WLAN where both classes have the same number of stations. The AP has $n$ downlink flows individually destined to each of the stations. There are no competing uplink flows.  
Flows are {\color{black}C}onstant {\color{black}B}
it {\color{black}R}ate (CBR) traffic with a fixed packet size of 1500 bytes. 
The queue size is 500 packets. 


%

Similarly to the theoretical performance analysis, we use experimental channel data shown in Fig.~\ref{outdoorchannelcapacity}.  We
assume that class 1 has a RSSI of 12dBm, and class 2 has a RSSI of
35dBm. The uncoded multi-destination aggregation approach uses a PHY rate of 18Mbps, while the coded
approach uses a PHY rate of 36Mbps (Note that this choice of PHY
rates is not necessarily optimal). The CBR traffic arrival rate is 1Mbps.

\begin{figure}
\centering \subfigure[Per downlink flow throughput]{
  \includegraphics[width=0.46\columnwidth, height=4.5cm]{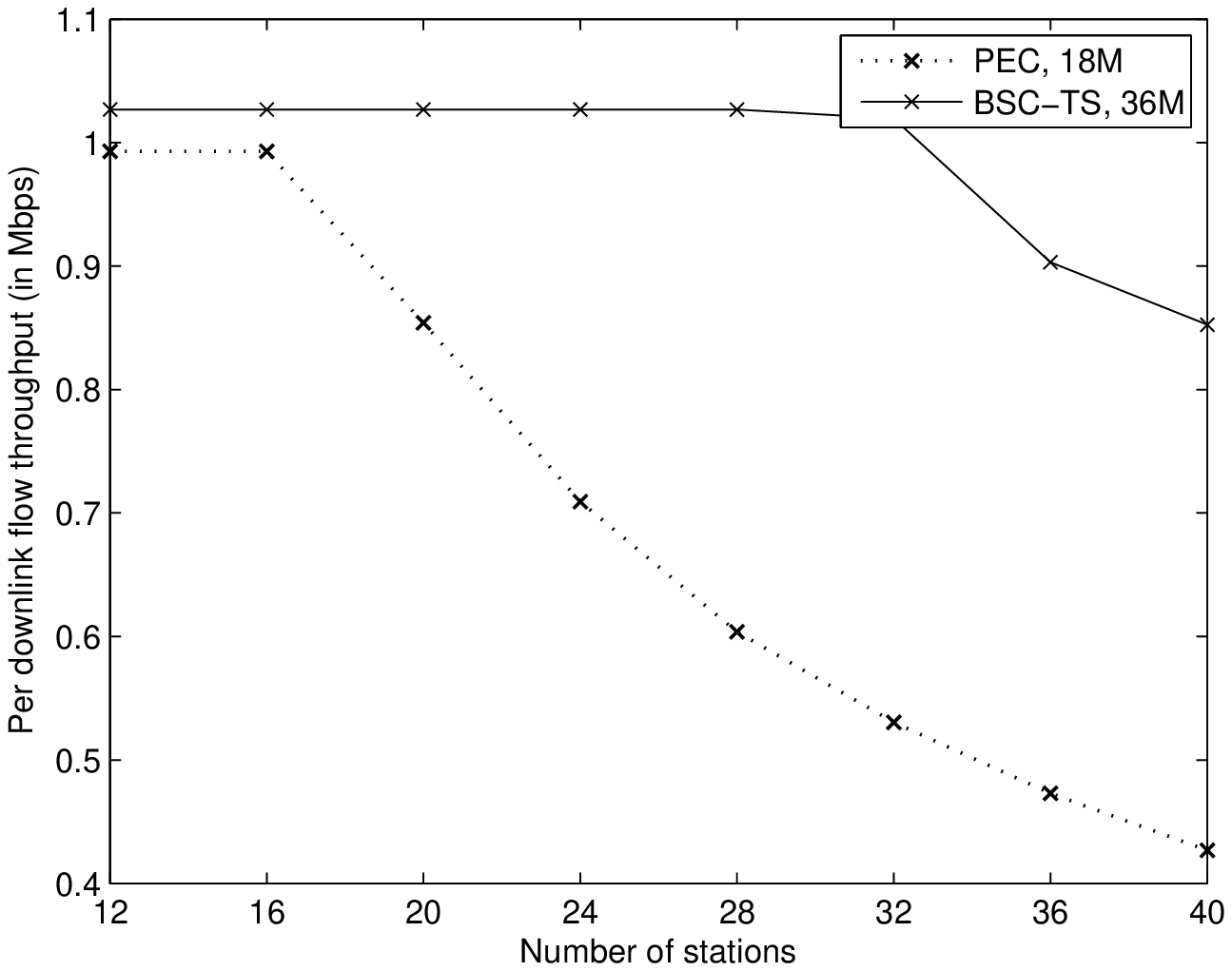}\label{exprthrpt}
 }
 \subfigure[Mean downlink delay]{
   \includegraphics[width=0.46\columnwidth, height=4.5cm]{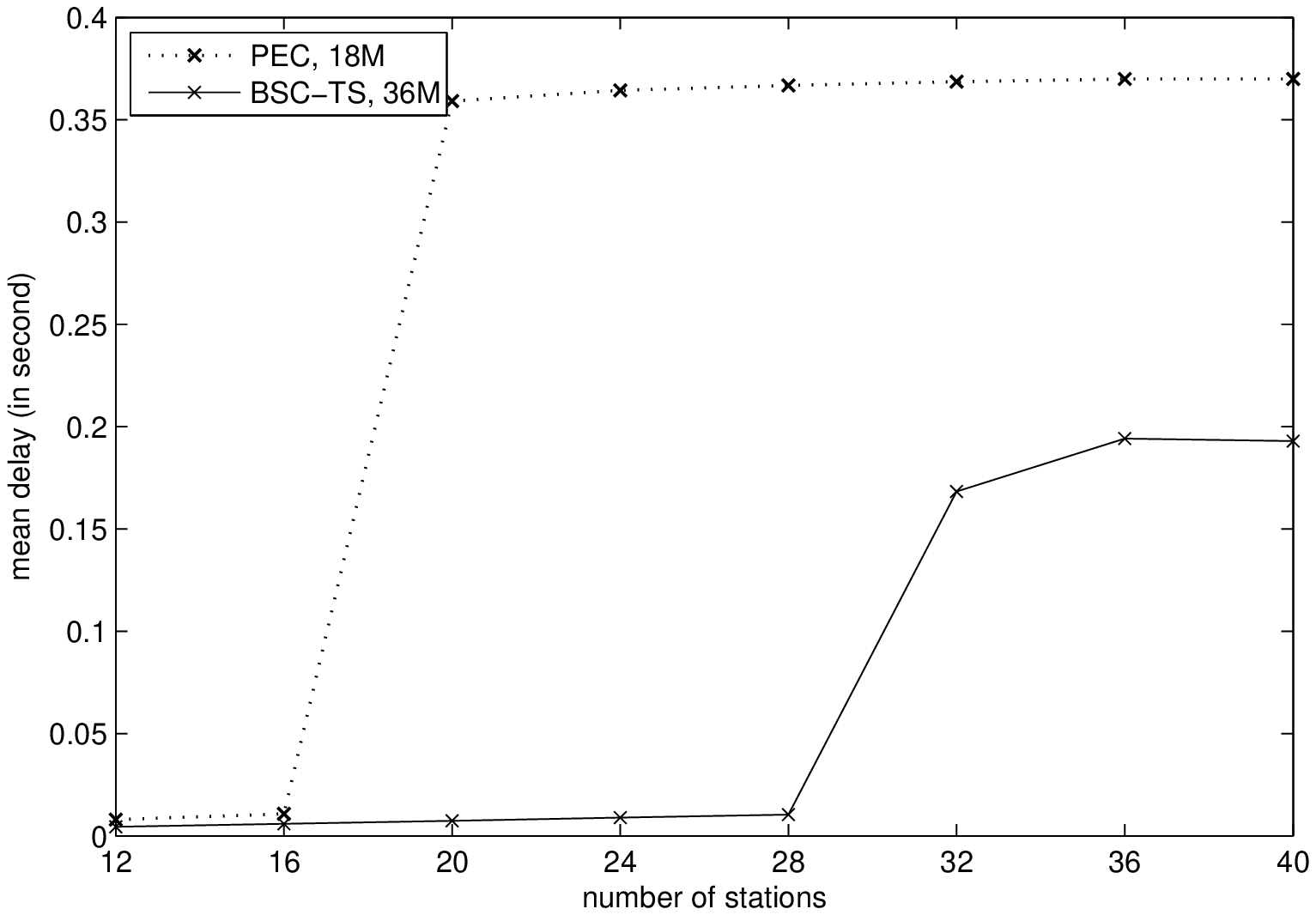}\label{exprdelay}
}
  \caption{Per downlink flow throughput and mean downlink delay vs number of stations, outdoor measurement channel data, RSSI of class 1 18dBm, RSSI of class 2 33dBm}\label{ExprExample}
\end{figure}

Fig.~\ref{ExprExample} plots the measured per downlink flow
throughput as the number is stations is varied. As expected, the
time-sharing coding scheme is strictly better than the uncoded
scheme, achieving higher throughput and lower delay.   For larger
numbers of stations, it can be seen that the time-sharing coding
scheme offers an almost 100$\%$ increase in per-flow throughput,
and the mean delay is half that of the uncoded scheme. When the
number of stations is below 28, the mean delay of time-sharing
scheme is extremely small. This is because when using the coded
scheme with such small numbers of flows, the queue in the AP is
mostly empty.  As the number of stations increases above 28, the
queue becomes backlogged and the mean delay (which includes the
packet waiting time in the queue) starts to increase.  In addition
the substantial increase in throughput and decrease in delay, we
also find that with time-sharing coding the AP can support
significantly more stations. It can be seen from
Fig.~\ref{exprdelay} that with uncoded multi-destination
aggregation the AP queue starts to become backlogged when the
number of stations rises above 16.  In comparison, with
time-sharing coding AP queue does not start to become backlogged
until the number of stations increases above 28.

\section{Discussions}

\subsection{Generalisation to a uniformly distributed error-prone WLAN }

The analytical work in Section~\ref{unicast},~\ref{multicast} can be generalised to a universal scenario where $n$ stations are uniformly distributed over the area in a WLAN, and each of them has an independent error-prone channel. In the PEC paradigm, a transmission from the AP fails ( i.e. AP doubles its contention window) only if none of the sub-frames is acknowledged by one of the multiple receivers. This could be caused by either a collision or noise-related erasures for all of the receivers.  Similarly, a transmission from an ordinary station fails also due to collisions or noise-caused erasures.  Thus, for both the AP and ordinary stations, the probability of a transmission fails is given by 
\begin{equation}\label{MACnew}
{p_f}_i=1-(1-{p_c}_i)(1-{p_e}_i), i=\{0, 1, 2, \cdots, n\}
\end{equation}
in which ${p_c}_i=1-\prod\limits_{{j=0, j\neq i}}^{n}(1-\tau_j)$ and ${p_e}_i=1-(1-{p_u}_i)^{L_i}$ with ${p_u}_i$ the first-event error probability in Viterbi decoding and $L_i$ the length of frame in bits. In the BSBC paradigm, since there are no noise-related packet erasures, packet losses are only caused by collisions, the transmission failure probability is ${p_f}_i={p_c}_ i$.

Apart from the difference in the MAC throughput model, the calculation of the expected payload for each flow and the expected MAC slot duration is similar.  If max-min fairness is considered, that is to equalise the flow throughput, following the same methodolody for our two-class running example,  the packet size for each downlink or uplink flow can be solved by combining the MAC model relationship Eqn.~\ref{18new} and Eqn.~\ref{MACnew} with the specific packet organisation requirement in each scheme. 

\subsection{Extension to other fairness criteria}

The proposed BSBC coding multi-destination aggregation schemes can be considered along with other fairness criteria, e.g. proportional fairness~\cite{PropFair}. The analysis will be established on a utility function in terms of the fairness requirement and specific constraints. An analytical or numeral solution to achieve the fairness objective can be obtained by using some optimisation method. The analysis for other fairness criteria is beyond the scope of this paper. To implement the proposed schemes in more general WLAN scenarios, we will consider different fairness criteria in the future.

\subsection{Implementation on standard hardware}
The present paper focuses on fundamental theoretical aspects. The experimental demonstration of a fully working system is out of scope.  We nevertheless comment briefly on the compatibility of the proposed coded multi-destination aggregation schemes with existing 802.11n hardware.   To implement  multi-destination aggregation with time-sharing coding on standard hardware, a fairly direct approach would be to aggregate MPDUs destined to different receivers into an A-MPDU frame.   Many 802.11 chipset drivers (e.g. atheros, broadcom) can be easily modified so as not to discard corrupted frames \emph{e.g.} see \cite{80211Hybrid}.    Encoding/decoding of the MPDU payload could then be carried out by a shim within the driver, and this would be transparent to higher network layers.   The 802.11 Block ACK functionality could be used to manage generation of MAC ACKs, or alternatively the 802.11 standard supports transmission of unicast packets with a ``No ACK'' flag set in the header and by using this ACKs could then be generated at a higher layer.   A less efficient user-space approximation to this scheme that requires no driver changes could be to encode packet payloads in user-space and use TXOP bursting to send these packets in a back-to-back burst (albeit with higher overhead than A-MPDU aggregation).  At the receiver, recent versions of the pcap API (or tcpdump) allow corrupted frames to be collected, where decoding could then take place in user-space. 
{\color{black}The channel state information (CSI) which is used for adaptive rate control at the physical layer needs to be passed upwards to the application layer for the AP to update the coding rate for each channel. }

\section{Conclusions}
In this paper we consider the potential benefits of viewing the
channel provided by an 802.11 WLAN  as a binary broadcast channel, as opposed to a
conventional packet erasure channel.   We propose two approaches
for multi-destination aggregation, \emph{i.e.} superposition coding and a simpler time-sharing
coding.   We develop throughput models for these coded
multi-destination aggregation schemes.   To our
knowledge, this provides the first detailed analysis of multi-user coding in 802.11 WLANs.
Performance analysis for both unicast and multicast traffic, taking account of
important MAC layer overheads such as contention time and
collision losses, demonstrate that increases in network throughput
of more than 100\% are possible over a wide range of channel
conditions and that the much simpler time-sharing scheme yields
most of these gains and have minimal loss of performance.
Importantly, these performance gains involve software rather than
hardware changes, and thus essentially come for ``free''.

\bibliographystyle{spmpsci}      
\bibliography{BSC_Multidest_ref}   



\end{document}